\documentstyle[eqsecnum,epsf,aps,prb]{revtex}

\begin{document}
\draft
\preprint{HEP/123-qed}

\twocolumn[\hsize\textwidth\columnwidth\hsize\csname
@twocolumnfalse\endcsname

\title{ Magnetic field - temperature phase diagram of quasi-two-dimensional organic superconductor $\lambda$-(BETS)$_2$GaCl$_4$ studied via thermal conductivity }

\author{ M. A. Tanatar$^{1,2,}$\cite{email,ikhp}, T. Ishiguro$^{1,2}$, H. Tanaka $^{3}$, and H. Kobayashi$^{3}$ }

\address{
$^1$ Department of Physics, Kyoto University, Kyoto 606-8502, Japan\\
$^2$ CREST, Japan Science and Technology Corporation, Kawaguchi, Saitama 332-0012, Japan\\
$^3$ Institute for Molecular Science, Myodaiji-cho, Okazaki 444-8585, Japan\\}

\date{\today}
\maketitle
\begin{abstract}
{
The thermal conductivity $\kappa$ of the quasi-two-dimensional (Q2D) organic superconductor $\lambda$-(BETS)$_2$GaCl$_4$ was studied in the magnetic field $H$ applied parallel to the Q2D plane. The phase diagram determined from this bulk measurement shows notable dependence on the sample quality. In dirty samples the upper critical field $H_{c2}$ is consistent with the Pauli paramagnetic limiting, and a sharp change is observed in $\kappa (H)$ at $H_{c2 \parallel}$. In contrast in clean samples $H_{c2}(T)$ shows no saturation towards low temperatures and the feature in $\kappa(H)$ is replaced by two slope changes reminiscent of second-order transitions. The peculiarity was observed below $\sim$0.33$T_c$ and disappeared on field inclination to the plane when the orbital suppression of superconductivity became dominant. This behavior is consistent with the formation of a superconducting state with spatially modulated order parameter in clean samples. 
}
\end{abstract}

]\narrowtext

\section{Introduction}

In most of superconductors, the suppression of superconductivity by a magnetic field $H$ is caused by the increase of diamagnetic energy. Since the effect is orbital in nature, the upper critical field $H_{c2}$ becomes high when the orbital motion is suppressed. Under this condition, the destruction of spin-singlet state of the Cooper pairs may become dominant mechanism of SC state suppression, with the $H_{c2}$ determined by the Pauli paramagnetic limiting field $H_p$ \cite{CC}. Both of these mechanisms give saturation in the temperature $T$ dependence of $H_{c2}$ at low temperatures \cite{Sarma,WHH}. In a number of recent experiments, however, a nonsaturating behavior of the $H_{c2}(T)$ was observed \cite{IshiguroJPhys}, and its possible relation to the formation of the inhomogeneous superconducting state was considered.

The existense of inhomogeneous state was predicted back in the sixties by Fulde and Ferrel \cite{FF} and Larkin and Ovchinnikov \cite{LO}, who pointed out that the stability of the superconducting (SC) phase can be increased above $H_p$ by pairing electrons with different momenta. In their model at low temperatures the transition from the normal state at $H_{ c2}$ proceeds into a new SC phase with non-zero momentum of the Cooper pair and spatially modulated order parameter, abbreviated recently as FFLO state, and becomes second order, contrary to the first order transition expected at $H_p$. Simultaneously at lower magnetic field an additional phase boundary appears within the SC domain, associated with the transition from FFLO state to usual SC state with zero momentum of the Cooper pair (hereafter we call it BCS state). 

Despite quite long history of theoretical studies \cite{GG,Aslamazov,Bulaevskii,Burkhard,BuzdinTugushev,Shimahara,Tachiki,MakiWon,Yang,AbrikosovFFLO,Klein}, the experimental observation of FFLO state is controversial. The state can be formed only if orbital motion is strongly suppressed, so that respective upper critical field $H_{c2 orb}$ becomes larger than $H_p$ \cite{conditions}, while the superconductor remains in a clean limit \cite{Aslamazov}. These conditions are not met simultaneously in usual superconductors, in which the orbital effect is suppressed by alloying for example. On the other hand, favourable situations for the formation of the FFLO state are found in heavy fermion superconductors \cite{Buzdin} and in quasi-two-dimensional (Q2D) superconductors under a magnetic field parallel to the Q2D plane \cite{Bulaevskii}; in these cases, due to the large effective mass, $H_{c2 orb}$ becomes very large even in clean samples. 

Indeed in several experimental studies, pronounced anomalies that were observed close to the $H_{c2}$ in Pauli paramagnetic limiting range in heavy-fermion \cite{Buzdin93,Modler,UBe,UBe2}, organic \cite{SingletonRes,SingletonMagn} and cuprate \cite{pulse} superconductors were discussed in a context of the formation of FFLO state. 
It should be noticed, however, that in all these studies one cannot rule out another effects appearing near $H_{ c2}$, especially in the resistivity $\rho$ and magnetization $M$ measurements. These can include anomalies related to the transformations in the vortex state \cite{Blatter}, such as peak effect \cite{peakeffect}, commensurability \cite{commensurability} and melting transition \cite{melting}, and the effects of filamentary and surface superconductivity \cite{surface} on measured $H_{c2}(T)$ line. Identification of the FFLO state through the shape of the $H_{c2}$ line with upturn at low temperature \cite{SingletonMagn,pulse} is not reliable as well, since the similar shape finds explanation in a number of scenarios \cite{LebedOld,Dupuis,LebedRestoration,Alexandrov,Abrikosov,Geshkenbein,Kresin}.  

A key property for the identification of FFLO state is its sensitivity to disorder. While the FFLO state should be rapidly suppressed by disorder, this is not true for the transformations in the vortex lattice. Indeed, alloying was found to increase anomalies close to $H_{c2}$ \cite{alloying}, attributed initially \cite{CeRuFFLO} to the FFLO state in CeRu$_2$. However, the effect of disorder is not unique for FFLO state, since it is similarly important \cite{Mineev} for several models explaining both high upper critical field values and upturn in the $H_{c2}(T)$ line on going to $T$=0.

In view of this it is desirable to make a more detailed experimental study on the subject. 
As useful technique to identify the FFLO state, we adopted the measurement of thermal conductivity $\kappa$. Its main advantages are bulk nature of the signal and insensitivity to the transformation in the vortex system. The former allows us to disregards filamentary and surface contributions, making dominating contributions to the resistive measurement \cite{MgB2}. The latter is related to the absence of Lorenz force acting on vortex lines, since the thermal current of quasi-particles (QP) does not create charge flow. As a consequence, the flux creep phenomena, determining $\rho$ and $M$, are not essential in $\kappa$, allowing determination of intrinsic $H_{c2}$ \cite{Hc2}. Within the mixed state, the electronic part of $\kappa$ in conventional SC is caused by QP tunnelling between vortices \cite{Sarma,Lowel}, while in unconventional SC the main contribution comes from the delocalized QP \cite{Volovik}. Both of these mechanisms are not influenced by the transformations of the vortex state and are not accompanied by any feature in $\kappa(H)$ \cite{features}. Therefore, the transition from the BCS state to FFLO state, which is expected to be accompanied by the release of QP  \cite{Tachiki}, should be directly seen as an increase in $\kappa(H)$. It is also advantageous that $\kappa$ can distinguish first order transition, leading to a jump in $\kappa(H)$ near $H_{c2}$ \cite{Suderow,CeCoIn}, from second order transition giving an anomaly in the slope of $\kappa(H)$ \cite{Sarma}.

In this article we report the study of thermal conductivity at low temperatures in the Q2D organic superconductor $\lambda$-(BETS)$_2$GaCl$_4$ in $H$ aligned precisely parallel to the conducting plane so as to suppress orbital motion. This superconductor is a non-magnetic version of $\lambda$-(BETS)$_2$FeCl$_4$ showing field-induced superconductivity due to a Jaccarino-Peter effect \cite{Uji,Balikas}. It was proposed theoretically that the FFLO state may be formed in both of these salts \cite{Balikas,BuBu}. This study became possible by notable technical advancement in thermal conductivity measurements in high and oriented magnetic fields \cite{Tanatar}. We show that the dependence of $H_{c2}$ on temperature, inclination angle and sample quality, together with sharpening of transition in $\kappa(H)$ curve with disorder, is consistent with the formation of FFLO state in clean samples of this organic superconductor.

\section{Experimental}

Measurements of thermal conductivity were performed using a standard steady state one-heater-two-thermometers technique. A sample was attached via annealed 10 $\mu$m Pt wires to a copper base of a miniature vacuum cell \cite{cell} and to RuO$_2$ chip resistors  \cite{RuO} acting as both heater and thermometers. The thermometers were calibrated in the temperature range 0.3 to 10 K at a set of magnetic fields and the magnetoresistance correction was made for in-field thermometry. The contacts to the sample were made by gold evaporation on fresh sample surface and subsequent wire attachment to a pad with {\it Dotite} silver paint. This technique allowed us to get contact resistance in the range of 100 to 500 m$\Omega$ at low temperatures. The same contacts were used for resistance and thermal conductivity measurements. The direction of the heat flow was along the longest dimension, corresponding to the crystallographic $c$ axis within conducting plane. 

The cell with the thermal conductivity measurement unit was rotated in $^3$He ambient by a double-axis goniometer in a superconducting solenoid.
The parallel alignment of a magnetic field to the Q2D plane of the sample was performed by measuring sample resistivity as a function of an inclination angle $\Theta$ in a magnetic field close to $H_{c2 \parallel}$ and determining the position of the minimum in $\rho (\Theta)$ curve. The accuracy of this alignment was determined by the sharpness of the feature and was typically $\pm $0.1$^{\circ}$.

Single crystals of $\lambda$-(BETS)$_2$GaCl$_4$ were grown by the electrochemical method \cite{crystals}. The samples had typical size of 2$\times$0.1$\times$0.1 mm$^3$. We studied 6 single crystals from 3 different batches, their properties are summarized in Table \ref{table}. A notable problem with measurements on these samples comes from resistance jumps on cooling. Despite use of soft Pt wire support and a precise control of the cooling rate at 0.5 K/min \cite{Kawasaki}, resistance jumps occurred in every second sample, especially in the range of structural transformation near 100 K, giving notable scatter in the residual resistivity value $\rho_0$ at temperatures just above the onset of the superconducting transition. The frequent jumps were to some extent caused by unfriendly stress conditions in thermal conductivity unit. Contrary to the resistivity, effect of the jumps was not pronounced in the thermal conductivity. The jumps are usually related to the formation of structural domain walls \cite{Ishiguro}, but this experiment indicates that the wall has different effect on charge and heat flow. The thermal transport is not influenced much by the boundary provided that the acoustic impedance is not changed substantially and therefore the break in electronic heat flow at a wall is healed by the phonon contribution. 

\section{Results}

At room temperature, the resistivity $\rho$ of the samples studied was in the range from 140 to 190 $\mu \Omega$m, with $\pm$ 10 \% error bar coming mainly from the uncertainty in the determination of the geometrical factor on thin samples with parallelogram cross-section. The temperature dependence of $\rho$ above 90 K was very similar among the samples from three batches studied, matching previous reports \cite{crystals,JSupercond}. On cooling, $\rho$ remained almost constant down to $\sim$200 K, increased slightly to $\sim$ 90 K and then decreased rapidly all the way down to the superconducting transition. The magnitude of the decrease below 90 K was notably dependent on sample batch, with variation in the residual resistance ratio from 20 to 150. In Fig. \ref{rt} we show $\rho(T)$ curves for two best samples showing no jumps on cooling. The samples belonged to batches A and B, characterized by a notably different residual resistance. The samples from batch C were of high quality as well, however, they were notably thinner and we have not succeeded in getting jump-free resistance curve for sample \#6. Worth of noting that $\rho(T)$ dependence just above the transition temperature $T_{\rm c}$ in the high $RRR$ sample is well described by  $T^2$-dependence ascribable to electron-electron ($ee$) scattering.

The $\kappa(T)/T$ in the SC region is shown in Fig. \ref{kappattd}. It increases on entering the SC state, and then shows the maximum at temperature $T_{max}$ with magnitude strongly dependent on sample quality. This peak is known in a wide range of superconductors \cite{Sarma}. It is caused by the condensation of normal electrons due to formation of a superconducting gap. In the normal state, conduction electrons are acting as scatterers for both heat carriers, electrons \cite{thermalHall} (via electron-electron ($ee$) scattering) and phonons (via phonon-electron ($ep$) scattering), determining their mean free paths, $l_e$ and $l_g$, respectively. On condensation the mean free path increases until reaching a limit determined by impurity and boundary scatterings, therefore the difference in the magnitude of the increase provides another way for characterisation of sample quality. The ratio of $\kappa/T$ at $T_c$ and at $T_{max}$ is shown in Table \ref{table}. The values are well reproduced between samples from the same batch. There was a clear correlation between the residual $\rho$ above $T_c$ (when this quantity was not influenced by resistance jumps) and the magnitude of $\kappa/T$ increase, as can be seen for the samples \#1 and \#2.

The $T_{c mid}$ was determined from the mid point giving 1/2 of $\rho_0$ evaluated at the junction point of the tangential lines for $\rho(T)$ in the normal and transition states. The derived values are approximately the same for all samples studied (see Table \ref{table}). The resistive transition is rather broad \cite{width}, but its width decreases systematically with the decrease in $\rho_0$. For sample \#1 the resistive $T_{c mid}$ appearing at 5.2 K corresponds to the onset of $\kappa /T$ increase. In contrast, the superconducting transition for \#2 did not give clear anomaly in $\kappa /T(T)$ at $T_c$, which makes the determination of $T_c$ and $H_{c2}$ from $\kappa$ measurements imprecise at high $T$. However, $\kappa(H)$ gives clear anomaly at $H_{c2}$ at low $T$, caused by a rapid change of $\kappa ^e$, as seen in both samples (Figs. \ref{flow},\ref{tcfd}), enabling high precision in $H_{c2}$ determination at low $T$.

In Fig. \ref{flow} we show $\kappa(T)$ at 0.3 K, the base temperature of our experiment. The dependence was taken as a function of $H$ of different orientation with respect to the heat flow and conducting plane in sample \#1. The $\kappa(H)$ dependence in the field applied perpendicular to Q2D plane is usual as for most of the superconductors at low temperatures. On the field increase, $\kappa$ decreases gradually then takes a minimum followed by the rapid increase towards $H_{c2}$. In the normal state above $H_{c2}$, $\kappa$ decreases with field due to the magnetoresistance effect on $\kappa_e$. The shape of the curve can be easily understood if we recall that $\kappa$ contains contributions of two heat carriers, electrons $\kappa_e$ and phonons $\kappa_g$, with $\kappa$ = $\kappa_e$ +$\kappa_g$. 
Since the density of phonons is not influenced by a magnetic field, the effect of magnetic field on $\kappa_g$ comes solely from the scattering. The phonons are effectively scattered by conduction electrons, and as we pointed out above, their mean free path is notably increased within the SC state. In the field the phonons become scattered by vortices. This effect is responsible for initial decrease of $\kappa$ under the perpendicular field, as shown in Fig. \ref{flow}. The decrease of $\kappa_g$ with $H$ is gradual and is well described by 1/$H$ law \cite{Lowel}. $\kappa_e$ gradually increases with field at low temperatures (both for conventional and unconventional superconductors) \cite{Lowel,Vekhter}, while it increases rapidly towards $H_{c2}$ in both cases due to the intervortex tunnelling \cite{Lowel,Tanatar}. Finally at $H_{c2}$ the increase of the electronic contribution is ceased and $\kappa_e$ decreases due to magnetoresistance in the normal state. 

Assuming the validity of the Wiedemann-Franz (WF) law, we can determine $\kappa_e$ from $\rho $ measured with the same geometrical factor when the same contacts are used. This estimation becomes evidently invalid when the sample shows resistance jumps, since weak links at domain walls behave differently for charge and heat transport. In the samples which showed no resistance jumps this estimation gives $\kappa_e$ equal to 30\% of total $\kappa$ in sample \#1 and 10\% in sample \#2. We show these values with double ended arrows in Figs. \ref{flow} and \ref{tcfd}. Another way to estimate the electronic contribution is to assume the same change of $\kappa$ and $\rho$ with magnetic field \cite{Roeshke}. This way of estimation produces the values which are quite close to the estimation via the WF ratio.

Since $\kappa_g$ is gradually decreasing with $H$ and all increase in $\kappa$ in the proximity to $H_{c2}$, which is of most interest for us, is electronic in origin, we consider phonon contribution as a background. The justification for this can be found in the dependence of $\kappa$ on the field orientation with respect to the heat flow direction. While the phonon conductivity is slightly varying with the field direction, the increase near $H_{c2}$ does not depend on it. The magnitude of increase in $\kappa$ near $H_{c2}$ can be regarded as a lower bound of the electronic thermal conductivity. As can be seen from Fig. \ref{flow} the magnitude of the increase is in quite reasonable correspondence with the estimation of $\kappa_e$ via the WF law.

In the parallel field, $\kappa(H)$ shows basically the same features as in the perpendicular field, except for several special points. The decrease of $\kappa_g$ is still gradual (as can be more easily seen in $\kappa(H)$ at higher temperatures, Fig. \ref{tcfd}). Due to the anisotropy of the coherence length, the cross-section of phonon scattering by vortices in the parallel field is notably smaller than in the perpendicular field \cite{phonon}, and the decrease of the phonon contribution is smaller, also. The cross-section of scattering by vortices slightly depends on the orientation of the field with respect to the thermal flow, as can be seen in Fig. \ref{flow}. However, $\kappa_e$ shows notably different behavior in the parallel field, as compared to the perpendicular field, and this difference is directly related to the Pauli limiting \cite{Suderow}. When the transition at $H_{c2}$ is caused by the orbital effect, vortices gradually fill the volume of the superconductor. On approaching $H_{c2}$ the distance between vortices gradually decreases, giving high probability of quasi-particle tunnelling between vortices. Because of this, $\kappa_e$ shows gradual and rapid increase near $H_{c2}$. In case of Pauli limiting the vortex matter does not fill all the volume of the superconductor at $H_{c2}$, which is a consequence of a first order transition at $H_p$. As a result, the change of $\kappa_e$ becomes stepwise, provided the system is clean. Actually the jump-like feature was observed in the Pauli paramagnetic limiting range in UPt$_3$ \cite{Suderow} and CeCoIn$_5$ \cite{CeCoIn}. 

It should be pointed out that the behavior of thermal conductivity in parallel field at low temperatures is well reproduced between different samples in the same batch and correlates with the magnitude of the peak in $\kappa (T)/T$. In Fig. \ref{samples} we show the field dependence of $\kappa$ in parallel field in the samples under study, and in Table \ref{table} we summarize the parameters of the $\kappa(H)$ curves at the base temperature. These parameters include the value of a field $H^*$ where $\kappa$ starts to increase, the magnitude of $\kappa$ increase from $H^*$ to $H_{c2}$, $\Delta \kappa$= $\kappa (H_{c2 \parallel})$- $\kappa (H^*)$, and the value of $H_{c2}$ itself. 

Worth of special note is the transformation of this behavior with disorder. In the low quality samples \#2 and \#5, $\kappa$ shows rather sharp increase in the proximity of $H_{c2}$, reminiscent of a broadened first order transition. In the high quality samples \#1,\#3,\#4 and \#6 the feature at $H_{c2}$ greatly broadens and is replaced by two slope changes, with the increase of $\kappa$ starting at lower and finishing at higher $H$ than in the dirty samples. 

To elucidate the origin of this behavior we studied its transformation with temperature and orientation of magnetic field both on inclination to conducting plane and on rotation parallel to it. Since these measurements are time consuming and the effect of the resistive jumps can not be completely ignored, we made the thorough studies on two best samples \#1 and \#2. In both cases the anomalous behavior near $H_{c2}$ disappears and $\kappa(H)$ becomes similar to the perpendicular field case for inclinations $\Theta$ above $\sim$5$^\circ$. 

In Fig. \ref{tcfd} we show $\kappa(H)$ for the field applied parallel to the plane ($\Theta$=0$^{\circ}$) at several $T$, with the curve for the perpendicular field $\Theta$=90$^{\circ}$ at 0.3 K, as a reference. As already pointed out, the anomalous behavior near $H_{c2}$ is observed only in the parallel field. (Fig. \ref{fdincl})

\section{Discussion}

As can be seen from Table \ref{table}, the behavior of thermal conductivity is consistent between the samples from the same batch, supporting that it is intrinsic. We will discuss the results by sticking to the behavior of two samples, for which we can get reliable value of $\rho$ at low temperatures due to the absence of resistance jumps. For these samples the difference in residual resistivity $\rho_0$ was the largest (Fig. \ref{rt}) with $\rho$=8$\times$10$^{-7}$ $\Omega$m for sample \#1 and 12$\times$10$^{-6}$ $\Omega$m for sample \#2. In standard conductivity theory this gives scattering time $\tau$ of 0.16 ps (\#1) and 0.01 ps (\#2). The $l_e$ can be estimated from the Fermi velocity averaged over the hole pocket of the Fermi surface, $v_F \approx 3*10^5$ m/s, \cite{Mielke} as 48 nm (\#1) and 3.2 nm (\#2).

In Fig.\ref{phase} we show the $H$ versus $T$ phase diagram for \#2 (coinciding within experimental accuracy to that of \#1) determined from $\rho(T)$ measurements together with the data determined from $\kappa(H)$ at low $T$. The slopes of the $H_{c2}(T)$ near $T_c$ for the perpendicular and parallel fields are equal to 0.45 T/K and 4.3 T/K, respectively. This gives the values of the coherence length $\xi_{\bot}(0)$=1.2 nm (out-of-plane) and $\xi_{\parallel}(0)$=11.6 nm (in-plane). The comparison of $\xi_{\bot}(0)$ (effective for $H_{\rm c2}$ for $\Theta$=0$^{\circ}$) with the $l_e$ shows that \#1 is in the clean limit, while sample \#2 is on the border between clean and dirty limits. 

Within the SC state the resistivity is determined by the vortex motion and the increase in $\rho$ starts below $H_{c2}$, as determined from $\kappa(H)$ in \#2 (Fig. \ref{fdincl}). However, in \#1 the increase of $\kappa$ at 0.3 K starts at $H$ giving $\rho$ = 0. This implies that the increase is not related to the vortex motion, but reflects the QP flow in the bulk. In the inset of Fig. \ref{phase} we show the $H_{c2}$ data determined from the $\kappa(H)$ measurement shown in Fig. \ref{tcfd} for 2 samples together with data points of the onset of $\kappa(H)$ increase in \#1.  It is noteworthy that in the high quality sample the linear increase of $H_{c2}$ on cooling correlates with the anomalous behavior of $\kappa(H)$. 

The anomaly observed in $\kappa (H)$ at $H_{c2}$ in the low quality sample \#2 is reminiscent of the behavior in the Pauli paramagnetic limiting range in UPt$_3$ and CeCoIn$_5$ \cite{Suderow,CeCoIn}, in which the first order transition at $H_{c2}$ leads to a jump in $\kappa$. Indeed, a clear saturation of $H_{c2}(T)$ at a value not far from the expectation for $H_p$ in the weak coupling BCS model ($H_p$=1.84$T_c$ with $H_p$ in T units and $T_c$ in K units), can be seen below $\sim$2 K (Fig. \ref{phase}). 
Taking $H_p$ as equal to the $H_{c2}$ of \#2 we can verify whether conditions for the formation of FFLO state are met in our experiment. With $H_{c2 orb}$=0.71$\frac{dH_{c2}}{dT}$ $\mid _{T_c}{T_c}$=16.3 T \cite{WHH} this leads to $\beta$(=$\frac{\sqrt{2} H_{c2o}}{H_p}$) =2.01 \cite{conditions}, which is suitable for the formation of FFLO state.

Although there is no jump in $\kappa(H)$ as well as hysteresis with field, which can confirm the first order type of transition directly, the transition is much sharper in dirty samples. On the contrary, in the high quality sample, $\kappa(H)$ in the region of $H_{c2}$ is broadened and replaced with two slope changes. This broadening in clean samples shows that some new process is involved. Indeed, it is natural to expect broadening of the phase transitions with disorder. This is really observed as broadening of the resistive transition near $T_c$ in zero field, but is strictly opposite to the behavior near $H_{c2}$. 
To characterize this feature we plot in the inset of Fig. \ref{phase} the field $H^*$, where the increase in $\kappa(H)$ starts in the high quality sample, as a function of $T$. Although the choice of this point at high $T$ is somewhat ambiguous due to steep phonon background variation, it is clear that $H^*(T)$ line extrapolates not far from the point where the $H_{c2}$ lines for the two samples begin to deviate, as expected for FFLO state. On the other hand, in sample \#2 the feature remains sharp, although due to much large contribution of $\kappa_g$, it is difficult to give any quantitative characterization of this fact. 

In Fig. \ref{angle} we show dependence of the $H_{c2}$ and of $H^*$ of sample \#1 on the orientation of the field within the two-dimensional plane. As it can be seen, none of the lines shows notable anisotropy, contrary to resistive measurements \cite{Kawasaki}, supporting their relation to Pauli paramagnetic limiting.

Summarizing our experimental observations we can state the following. (1) The shape of the phase diagram in the parallel magnetic field, $H_{c2 \parallel}(T)$, depends on the mean free path of samples. In clean samples the line shows no saturation at low temperatures, while it shows clear saturation in dirty samples. (2) The increase of $\kappa$ near $H_{c2 \parallel}$ gets essentially sharper in  dirty samples indicating a tendency towards the first-order transition, while the second-order transition is observed in clean samples. (3) The increase of $\kappa(H)$ above phonon background starts at lower fields in clean samples, indicating an additional phase boundary within the SC phase at low temperatures and high fields. (4) The difference is clearly observed below $\sim$1.7 K, i.e. $\sim0.33T_c$. (5) The anomalies are specific to the parallel field direction and disappear on the field inclination beyond $\Theta_c \sim 5^{\circ}$, where the orbital motion becomes dominating.

All these features are in line with theoretical predictions for the FFLO state. An important issue, however, is whether this explanation is unique. 
Several models were put forward \cite{Alexandrov,Abrikosov,Geshkenbein,Kresin} to explain an unusual upturn of $H_{c2}(T)$ at low $T$ in high-$T_c$ cuprates \cite{MacKenzie}. These are, however, not specific for the parallel field orientation. Among the models based on the low-dimensionality of the system, the most relevant is the model by Lebed and Yamaji (LY) \cite{LebedRestoration}. This model is a generalization for Q2D case of the original idea, developed by Lebed for Q1D systems  \cite{LebedOld}, and studied later as well by Dupuis, Montambaux and Sa de Melo (DMS) \cite{Dupuis}. We would refer to the LY theory, since its dimensionality is in accordance with the present case. The model takes into account quantum corrections to the orbital motion, becoming important if the length scales of the orbital confinement (interlayer distance $d$) and of the superconductivity ($\xi_{\bot}$) are comparable. The FFLO and LY theories make similar predictions with respect to the suppression of upturn of the $H_{c2}$ on inclination and on increase of impurity scattering \cite{Mineev}. There is, however, notable quantitative difference. An upturn of $H_{c2}(T)$ in LY theory starts below $T^*\approx 0.1(d/\xi_{\bot}(0))T_c(0)$. In $\lambda$-(BETS)$_2$GaCl$_4$ this gives $T^*\approx$ 0.6 K, notably lower than $\sim$1.7 K. In the FFLO state the upturn is predicted \cite{GG} to start at $T_i$=0.56$T_c$,  $\sim$2.7 K, if the orbital motion is negligible, which is 1.5 times higher than extrapolated in our experiment. The account of the orbital effect, however, is expected to shift the transition point to lower $T$, improving matching with the experiment in the present case. 

In addition, the LY, Lebed and DMS models do not break the paramagnetic limit $H_p$, therefore the sharpening of the anomaly in $\kappa(H)$ in dirty samples with lower $H_{c2}$ does not find natural explanation. Another apparent difficulty in the applicability of the LY theory is the existence of an additional phase boundary within the SC state, on which the increase of $\kappa$ starts in clean samples. Yet, this point may be not so clear, since a cascade of phase transitions is predicted in Q1D case in DMS model \cite{Dupuis}, some of which may survive in Q2D.  From the above consideration, we see that FFLO theory gives a better description of our experiment, both in respect of the onset temperature and the tendency for the first-order transition in dirty samples. 

As it can be seen none of the features of our data contradicts formation of the FFLO state. For identification of this state more clear observation of the first order transition in dirty samples may be useful \cite{BuBu}, however, final judgement can be made based on either phase-sensitive experiments \cite{Agterberg}, or direct observation of the structure of the order parameter in STM experiment \cite{STM}. 

\section{Conclusion}

In conclusion, the thermal conductivity in the organic superconductor $\lambda$-(BETS)$_2$GaCl$_4$ in the magnetic field applied parallel to the Q2D plane shows a clear anomaly of the field dependence near $H_{c2}$, correlated with the shape of the $H_{c2 \parallel}$ phase diagram at low temperatures. Both features disappear with the temperature increase and field inclination to the plane.
This is consistent with the Pauli paramagnetic limiting in the low quality samples, while with the FFLO state formation in the high quality samples. This observation may be of importance for explanation of the unusual field-induced superconductivity in the closely related $\lambda$-(BETS)$_2$FeCl$_4$ salt \cite{Uji,Balikas}.

\section{Acknowledgement}

The authors acknowledge A. G. Lebed, V. P. Mineev, H. Shimahara, Y. Maeno, M. Lang, and M.V. Kartsovnik for valuable discussions.

\begin{table*}
\caption{ The properties of 6 single crystals of $\lambda$-(BETS)$_2$GaCl$_4$. Samples showing resistance jumps are marked with *.}
\label{table}
\begin{tabular}{ccccccccc}
Sample No.      & \# 1  & \# 2  & \# 3 & \# 4 &  \# 5 & \# 6\\
batch & A & B  & A & A & B & C\\
\hline
     $\rho_{RT}$ ($\mu \Omega$m)  & 148 & 190  & 142  & 170 & 153 & 140 \\
     $\rho_{RT}/\rho_0$  & 150 &  20  & 23*  & 3.5* & 1* & 130* \\
     $T_{c,mid}$ (K)  & 5.2 & 5.6 &  5.5  & 5.1 & 5.2 & 5.1 \\  
     $\Delta T_c$ (K)  & 1.7 & 2.2 & 1.8  & 1.6 & 2.3 & 1.5 \\  
     $\frac{(\kappa /T)_{(T_{max})}}{ (\kappa/T)_{(T_c)}}$ & 1.35 & 1.01 &  1.38 & 1.30 & 1.06 & 1.40 \\
     $H_{c2 \parallel}(0.3 K)$ (T) & 12.7 & 11.2 &  12.9 & 12.6 & 11.1 & 13.3 \\
     $H^* (0.3 K)$ (T)  & 9.7 & 10.5 &  9.7 & 9.8 & 10.7 & 9.5\\

     $\Delta \kappa /\kappa _{H_{c2}}$ (\%)& 24 & 6 &  27  & 32 & 7 & 34 \\
\end{tabular}
\end{table*}

\begin{figure}
\begin{center}
	\epsfxsize=5cm
	\epsfbox{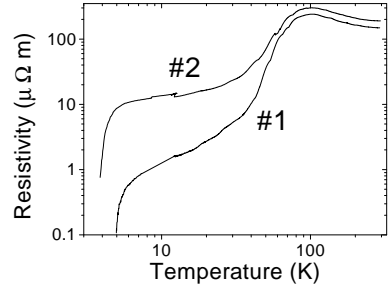}
\end{center}
\vspace{20mm}
\caption{
Temperature dependence of the resistivity for $\lambda$-(BETS)$_2$GaCl$_4$, samples \#1 and \#2. 
}\label{rt} 
\end{figure}

\vspace{10mm}
\begin{figure}
\begin{center}
	\epsfxsize=5cm
	\epsfbox{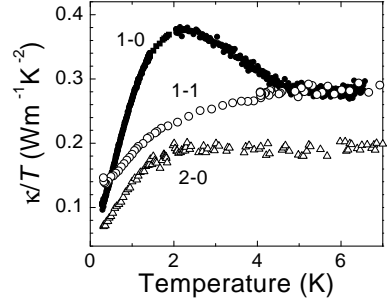}
\end{center}
\vspace{10mm}
\caption{
Temperature dependence of $\kappa/T$ for $\lambda$-(BETS)$_2$GaCl$_4$. Sample \#1 at $H$=0 (curve 1-0) and in the normal state (1-1) ($H$= 6 T applied perpendicularly to the Q2D plane); sample \#2 at $H$=0 (curve 2-0). 
}\label{kappattd} 
\end{figure}

\newpage
\begin{figure}
\begin{center}
	\epsfxsize=6cm
	\epsfbox{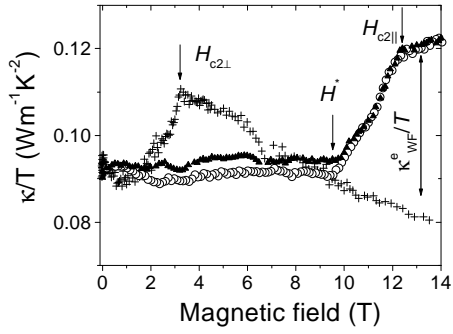}
\end{center}
\vspace{5mm}
\caption{
Field dependence of $\kappa/T$ for sample \#1 at base temperature of 0.3 K in a magnetic field perpendicular to the plane (along $b'$ direction, crosses), and parallel to the plane along (triangles)  and perpendicular (circles) to the heat flow direction ($c$ axis). Double-ended arrow shows estimation of $\kappa_e /T$ from Wiedemann-Franz ratio.
}\label{flow} 
\end{figure}

\newpage
\begin{figure}
\begin{center}
	\epsfxsize=5cm
	\epsfbox{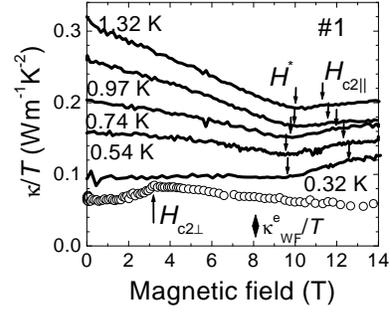}
\vspace{15mm}
\end{center}
\begin{center}
	\epsfxsize=5cm
	\epsfbox{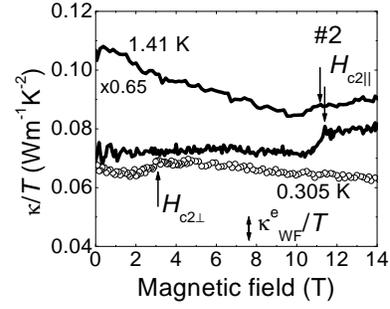}
\end{center}
\vspace{10mm}
\caption{
Field dependence of $\kappa/T$ for samples \#1 (a) and \#2 (b). The data under the field parallel to the Q2D plane at several temperatures are shown by solid lines. The data for the perpendicular field at 0.3 K (shown with open circles) are shifted downward to avoid overlapping. 
}\label{tcfd} 
\end{figure}

\newpage
\begin{figure}
\begin{center}
	\epsfxsize=6cm
	\epsfbox{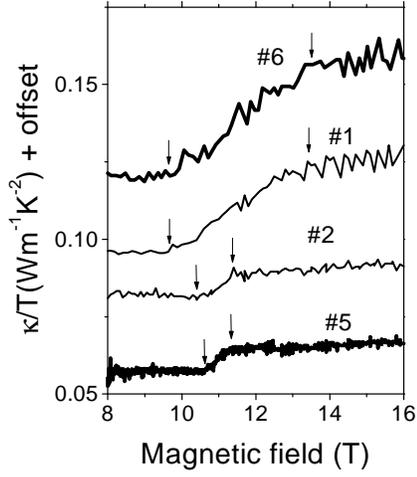}
\end{center}
\vspace{10mm}
\caption{
Field dependence of $\kappa/T$ for high quality samples \#1 and \#6 and low quality samples \#2 and \#5. The data were taken under the field parallel to the Q2D plane at base temperature 0.3 K. The curve for \#6 is shifted upwards by 0.04 Wm$^{-1}$K$^{-2}$ and for \#5 by the same amount downwards to avoid overlapping, the curves for \#3 and \#4 are similar to that for \#1 and are not shown. 
}\label{samples} 
\end{figure}

\newpage
\begin{figure}
\begin{center}
	\epsfxsize=6cm
	\epsfbox{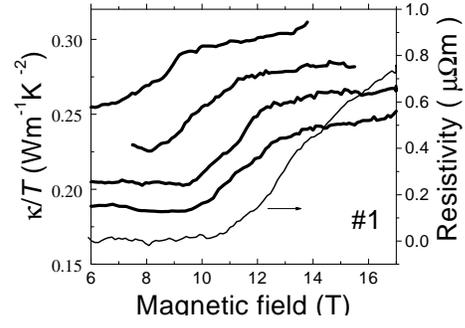}
\end{center}
\vspace{10mm}
\begin{center}
	\epsfxsize=6cm
	\epsfbox{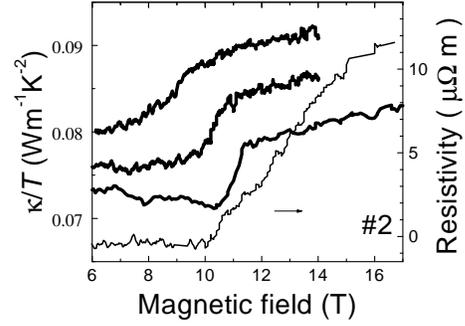}
\vspace{10mm}
\end{center}

\caption{
Field dependence of $\rho$ for $\Theta=0$ (thin line) and of $\kappa$ under the fields inclined to the Q2D plane (thick lines, shifted upward with $\Theta$ to avoid overlapping). $T$=0.3 K. (a) Sample \#1, $\Theta$= 0$^\circ$, 1$^\circ$, 2 $^\circ$ and 7$^\circ$ (from bottom to top). (b) Sample \#2, $\Theta$= 0$^\circ$, 1$^\circ$ and 5 $^\circ$. 
}\label{fdincl} 
\end{figure}

\newpage
\begin{figure}

\begin{center}
	\epsfxsize=5cm
	\epsfbox{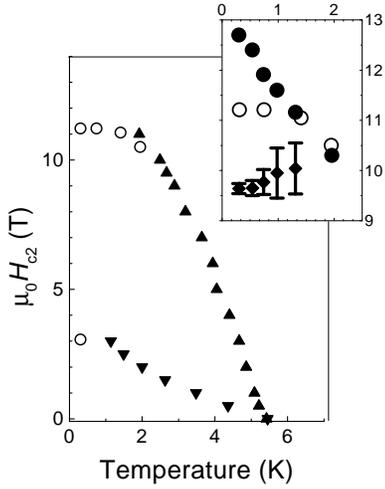}
\end{center}
\vspace{10mm}
\caption{
Phase diagram for sample \#2 determined from the midpoint of the resistive transition at fixed $T$ (up- and down-triangles correspond to parallel and perpendicular field orientations) and from the slope change in $\kappa (H)$ at fixed $T$ (open circles).  The inset shows the low-temperature part of the phase diagram determined from $\kappa(H)$: solid and open circles are $H_{c2 \parallel}$ for \#1 and \#2, respectively and diamonds correspond to $H^*$ (see text) of \#1. 
}\label{phase} 
\end{figure}

\newpage
\begin{figure}
\begin{center}
	\epsfxsize=6cm
	\epsfbox{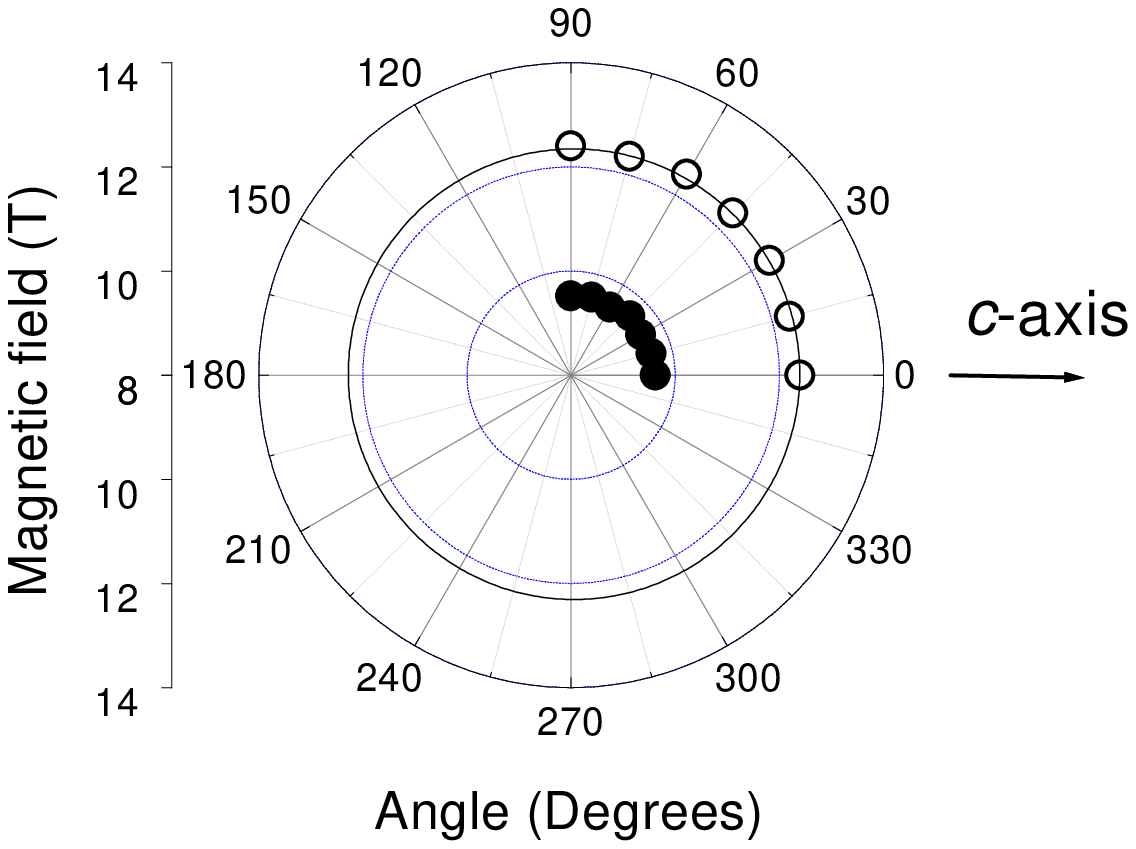}
\end{center}
\vspace{10mm}
\caption{
Dependence of fields $H_{c2 \parallel}$ (circles) and $H^*$ (triangles) for \#1 on the orientation of magnetic field within the conducting plane. The scale of the magnetic field is represented in the lefthand side in units of tesla. The size of the symbol corresponds to error bar. The solid line is a guide for the eye.
}\label{angle} 
\end{figure}


\begin{references} 
\bibitem[*]{email}Corresponding author, E-mail: tanatar@physics.utoronto.ca
\bibitem[\dagger]{ikhp} Present address: Physics Department, University of Toronto, Canada; Permanent address: Inst. Surface Chemistry, N.A.S. Ukraine, Kyiv, Ukraine. 

\bibitem{CC} A. M. Clogston, Phys. Rev. Lett. {\bf 9}, 266 (1962); B. S. Chandrasekhar, Appl. Phys. Lett. {\bf 1}, 7 (1962).

\bibitem{Sarma} D. Saint-James, G. Sarma, and E. J. Tomas, {\it Type II 
Superconductivity} Pergamon, Oxford, 1969. 

\bibitem{IshiguroJPhys} See T. Ishiguro, J. de Physique IV {\bf 10}, 139 (2000) for a review of the experimental data. 

\bibitem{WHH} N. R. Werthamer, E. Helfand, and P. C. Hohenberg, Phys. Rev. {\bf 147}, 295 (1966).

\bibitem{FF}
P. Fulde and R. A. Ferrell, Phys. Rev. {\bf 135}, A550 (1964).

\bibitem{LO}
A. I. Larkin and Yu. N. Ovchinikov, Zh. Eksp. Teor. Fiz. {\bf 47}, 1136 (1964) [Sov. Phys. JETP, {\bf 20}, 762 (1965)].


\bibitem{GG} L. G. Gutenberg and L. Gunter, Phys. Rev. Lett. {\bf 16} 996 (1966).
\bibitem{Aslamazov} L. G. Aslamazov, Zh. Eksp. Teor. Fiz. {\bf 55}, 1477 (1968) [Sov. Phys. JETP {\bf 28}, 773 (1969)]. 

\bibitem{Bulaevskii} L. N. Bulaevskii, Zh. Eksp. Teor. Fiz. {\bf 65}, 1278 (1973) [Sov. Phys. JETP {\bf 38} 634 (1974)]. 

\bibitem{Burkhard} H. Burkhardt and D. Rainer, Ann. Phys. {\bf 3}, 181 (1994).

\bibitem{BuzdinTugushev} A. I. Buzdin and V. V. Tugushev, Zh. Eksp. Teor. Fiz. {\bf 85}, 735 (1983) [Sov. Phys. JETP {\bf 58}, 428 (1983)].


\bibitem{Shimahara} H. Shimahara, J. Phys. Soc. Jpn. {\bf 67} 1872 (1997) and Phys. Rev. B{\bf 50}, 12760 (1994).

\bibitem{Tachiki} M. Tachiki, S. Takahashi, P. Gegenwart, M. Weiden, M. Lang, C. Geibel, F. Steglich, R. Modler, C. Paulsen, and Y. Onuki, Z. Phys. B. {\bf 100}, 369 (1996). 
\bibitem{MakiWon} K. Maki and H. Won, Czech. J. Phys. {\bf 46}, 1035 (1996).
\bibitem{Yang} K. Yang and S. L. Sondhi, Phys. Rev. B {\bf 57}, 8566 ( 1998).


\bibitem{AbrikosovFFLO} A. A. Abrikosov and A. I. Buzdin, Phys. Rev. B {\bf 63}, 224506 (2001).

\bibitem{Klein} S. Manalo and U. Klein, J. Phys.: Cond. Mat. {\bf 12}, L471 (2000).


\bibitem{conditions} The conditions for observation of FFLO state require that Maki parameter [K. Maki, Physics {\bf 1}, 127 (1964)] $\beta$ describing the ratio of orbital and paramagnetic effects to be larger than 1.8.

\bibitem{Buzdin} A. I. Buzdin and J.P. Brison, Europhys. Lett. {\bf 35} 707 (1996).

\bibitem{Buzdin93} K. Gloos, R. Modler, H. Schimanski, C. D. Bredl, C. Geibel, F. Steglich, A. I. Buzdin, N. Sato, and T. Komatsubara, Phys. Rev. Lett. {\bf 70}, 501 (1993).

\bibitem{Modler} R. Modler, P. Gegenwart, M. Lang, M. Deppe, M. Weiden, T. Luhmann, C. Geibel, F. Steglich, C. Paulsen, J. L. Tholence, N. Sato, T. Komatsubara, Y. Onuki, M. Tachiki, and S. Takahashi, Phys. Rev. Lett. {\bf 76}, 1292 (1996).

\bibitem{UBe} F. Thomas, B. Wand, T. Luhmann, P. Gegenwart, G. R. Stewat, F. Steglich, J. P. Brison, A. Buzdin, L. Glemot, and J. Floquet, J. Low Temp. Phys. {\bf 102}, 117 (1996).

\bibitem{UBe2} L. Glemot, J. P. Brison, J. Flouquet, A. I. Buzdin, I. Sheikin, D. Jaccard, C. Thessieu, and F. Thomas, Phys. Rev. Lett. {\bf 82}, 169 (1999).



\bibitem{SingletonRes} M-S Nam, J. A. Symington, J. Singleton, S. J. Blundell, A. Ardavan, J. A. A. J. Perenboom, M. Kurmoo, and P. Day, J. Phys.: Cond. Mat. {\bf 11}, L477 (1999).

\bibitem{SingletonMagn} J. Singleton, J. A. Symington, M-S Nam, A. Ardavan, M. Kurmoo, and P. Day, J. Phys.: Cond. Mat. {\bf 12}, L641 (2000).

\bibitem{pulse} J. L. O'Brien, H. Nakagawa, A. S. Dzurak, R. G. Clark, B. E. Kane, N. E. Lumpkin, R. P. Starrett, N. Muira, E. E. Mitchell, J. D. Goettee, D. G. Rickel, and  J. S. Brooks, Phys. Rev. B {\bf 61}, 1584 (2000).


\bibitem{Blatter} G. Blatter, M. V. Feigel'man, V. B. Geshkenbein, A. I. Larkin, and V. M. Vinokur, Rev. Mod. Phys. {\bf 66}, 1125 (1994).

\bibitem{peakeffect} See for example A. B. Pippard, Philos. Mag. {\bf 19} 217 (1969).
\bibitem{commensurability} A. A. Zhukov, M. G. Mikheev, V. I. Voronkova, K. I. Kugel', A. L. Rakhmanov, H. Kupfer, T. Wolf, G. K. Perkins, and A. D. Caplin
JETP Lett. {\bf 69}, 881 (1999). 

\bibitem{melting} E. Zeldov, D. Majer, M. Konzykowski, V. B. Geshkenbein, V. M. Vinokur, and H. Shtrikman, Nature, {\bf 375}, 373 (1995).

\bibitem{surface} A. Buzdin and T. Chameeva, Phys. Lett. A {\bf 207}, 113 (1995).

\bibitem{LebedOld} A.G.Lebed', Pis'ma Zh. Eksp. Teor. Phys. {\bf 44}, 89 (1986) [JETP Lett. {\bf 44}, 144 (1986)].
\bibitem{Dupuis} N. Dupuis, G. Montambaux, and C.A.R. S\'{a} de Melo, Phys. Rev. Lett. {\bf 70} 2613 (1993).

\bibitem{LebedRestoration} A. G. Lebed and K. Yamaji, Phys. Rev. Lett. {\bf 80}, 2697 (1998).
\bibitem{Alexandrov} A. S. Alexandrov, V. N. Zavaritsky, W. Y. Liang, and P. L. Nevsky, Phys. Rev. Lett. {\bf 76}, 983 (1996).
\bibitem{Abrikosov} A. A. Abrikosov, Phys. Rev. B {\bf 56}, 5112 (1997).
\bibitem{Geshkenbein} V. B. Geshkenbein, L. B. Ioffe, and A. J. Millis, Phys. Rev. Lett. {\bf 80}, 5778 (1998).
\bibitem{Kresin} Y. N. Ovchinnikov and V. Z. Kresin, Phys. Rev. B {\bf 54}, 1251 (1996).

\bibitem{alloying} N.R. Dilley and M. B. Maple, Physica C {\bf 278}, 207 (1997). 

\bibitem{CeRuFFLO} A. D. Huxley, C. Paulsen, O. Laborde, J.L. Tholence, D. Sanches, A. Junod, and R. Calemczuk, J. Phys.: Cond. Mat. {\bf 5}, 7709 (1993).

\bibitem{Mineev} V. P. Mineev, J. Phys. Soc. Jpn. {\bf 69}, 3371 (2000).

\bibitem{MgB2} A. V. Sologubenko, J. Jun, S. M. Kazakov, J. Karpinski, and H. R. Ott, Phys. Rev. B {\bf 65}, 180505 (2002).

\bibitem{Hc2} S. Belin, T. Shibauchi, K. Behnia, and T. Tamegai, J. Supercond. {\bf 12}, 497 (1999).

\bibitem{Lowel} J. Lowell and J. B. Sousa, J. Low Temp. Phys. {\bf 3}, 65 (1970). 
\bibitem{Volovik} G. E. Volovik, Pis'ma Zh. Eksp. Teor. Fiz. {\bf 58}, 457 (1993) [JETP Lett. {\bf 58}, 469 (1993)].

\bibitem{features} This can be inferred from comparison of magnetization and thermal conductivity measurements. Both data sets are available for UPt$_3$ with $\kappa$ of Ref.31 and magnetization by S.Schottl, E. A.  Schuberth, K. Flachbart, J. B. Kycia, W. P. Halperin, 
A. A. Menovsky, E. Bucher, and J. Hufnagl, Phys. Rev. B {\bf 62}, 4124 (2000), and for Sr$_2$RuO$_4$, H. Yaguchi, K. Deguchi, M. A. Tanatar, Y. Maeno, and T. Ishiguro, J. Phys. Chem. Solids, to be published. 

\bibitem{Suderow} H. Suderow, J. P. Brison, A. Huxley, and J. Flouquet, 
J. Low Temp. Phys. {\bf 108}, 11 (1997).

\bibitem{CeCoIn} K. Izawa, H. Yamaguchi, Y. Matsuda, H. Shishido, R. Settai, and Y. Onuki, Phys. Rev. Lett. {\bf 87}, 057002 (2001). 

\bibitem{Uji} S. Uji, H. Shinagawa, T. Terashima, T. Yakabe, Y. Terai, M. Tokumoto, A. Kobayashi, H. Tanaka, and H. Kobayashi, Nature {\bf 410}, 908 (2001).
\bibitem{Balikas} L. Balicas, J. S. Brooks, K. Storr, S. Uji, M. Tokumoto, H. Tanaka, H. Kobayashi, A. Kobayashi, V. Barzykin, and L. P. Gor'kov,
Phys. Rev. Lett. {\bf 87}, 067002 (2001).
 \bibitem{BuBu} M. Houzet, A. Buzdin, L. Bulaevskii, and M. Maley, Phys. Rev. Lett. {\bf 88}, 227001 (2002).

\bibitem{Tanatar} M. A. Tanatar, S. Nagai, Z. Q. Mao, Y. Maeno, and T. Ishuguro, Phys. Rev. B {\bf 63}, 064505 (2001).

\bibitem{cell} V. A. Bondarenko, M. A. Tanatar, A. E. Kovalev, T. Ishiguro, S. Kagoshima, and S.Uji, Rev. Sci. Instr. {\bf 71}, 3148 (2000).

\bibitem{RuO} Model RK73K1EJ manufactured by KOA.

\bibitem{crystals} H. Kobayashi, H. Tomita, T. Udagawa, T. Naito, and A. Kobayashi, Synth. Met. {\bf 70}, 867 (1995).

\bibitem{Kawasaki} T. Kawasaki, M. A. Tanatar, T. Ishiguro, H. Tanaka, A. Kobayashi, and H. Kobayashi, Synth. Met. {\bf 120}, 771 (2001).

\bibitem{JSupercond} M. A. Tanatar, T. Ishiguro, H. Tanaka, A. Kobayashi, and H. Kobayashi, J. Supercond. {\bf 12}, 511 (1999).

\bibitem{Ishiguro} T. Ishiguro, K. Yamaji, and G. Saito, {\it Organic Superconductors}, 2nd ed., Vol. 88 of Springer Series in Solid-State Physics (Springer, Heidelberg, 1998).
\bibitem{thermalHall} K. Krishana, J. M. Harris, and N. P. Ong, 
Phys. Rev. Lett. {\bf 75}, 3529 (1995).


\bibitem{width} This width is typical for organic superconductors, and is likely intrinsic in origin, see, J. Singleton, N. Harrison, C. H. Mielke, J. A. Schlueter, and A. M. Kini, J. Phys.: Cond. Mat. {\bf C 13}, L899 (2001).

\bibitem{Vekhter} I. Vekhter and A. Houghton, Phys. Rev. Lett. {\bf 83}, 4626 (1999).

\bibitem{Roeshke} F. Roeske, H. R. Shanks, and D. K. Finnemore, Phys. Rev. B {\bf 16}, 3929 (1977).

\bibitem{phonon} M. A. Tanatar, M. Suzuki, S. Nagai, Z. Q. Mao, Y. Maeno, and T. Ishiguro, Phys. Rev. Lett. {\bf 86}, 2649 (2001); M. Suzuki, M. A. Tanatar, Z. Q. Mao, Y. Maeno, and T. Ishiguro, Condmat 0104493 (2001).

\bibitem{Mielke} C. Mielke, J. Singleton, M. S. Nam, N. Harrison, C. C. Agosta, B. Fravel, and L. K. Montgomery, J. Phys.: Cond. Mat.  {\bf C 13}, 8325 (2001).


\bibitem{MacKenzie} A. P. Mackenzie, S. R. Julian, G. G. Lonzarich, A. Carrington, S. D. Hughes, R. S. Liu, and D. C. Sinclair, 
Phys. Rev. Lett. {\bf 71}, 1238 (1993).
\bibitem{Agterberg} K. Yang and D. F. Agterberg, Phys. Rev. Lett. {\bf 84} 4970 (2000).
\bibitem{STM} H. Shimahara, J. Supercond. {\bf 12}, 469 (1999).


\end{references}
\end{document}